%% file: RelativisticCVCSs.tex
\documentclass[prl,letterpaper,nofootinbib,superscriptaddress,twocolumn,notitlepage]{revtex4}

\input{macros}

\begin{document}

\title{Comment on ``Towards universal quantum computation through relativistic motion''}
\author{Rafael N. Alexander}
\author{Natasha C. Gabay}
\author{Nicolas C. Menicucci}
\affiliation{School of Physics, The University of Sydney, Sydney, NSW 2006, Australia}

\date{\today}

\begin{abstract}
The work reported in %
arXiv:1311.5619v1 proposes to produce continuous-variable cluster states through relativistic motion of cavities%
. This proposal does not produce the states claimed by the authors. %
The states actually produced are in general not known to be useful for measurement-based quantum computation. %
\end{abstract}

\maketitle

\emph{Inroduction.}---%
Ref.~\cite{Bruschi:2013uv} proposes a way to generate continuous-variable cluster states (CVCSs) through relativistic motion of an optical cavity. The proposal is to rapidly accelerate and decelerate the cavity multiple times, thus generating sizable two-mode squeezing~(TMS) interactions between particular cavity modes. (Alternatively, one can rapidly modulate the boundary conditions of a fixed cavity to achieve the same effect.) This is a fascinating suggestion since the usual way to generate TMS interactions is with nondegenerate parametric downconversion using a nonlinear optical crystal---in this proposal, the crystal has been entirely replaced with just relativistic motion. Unfortunately, we show in the following that the method prescribed in Ref.~\cite{Bruschi:2013uv} does not in general produce a CVCS with the claimed structure. %
We illustrate the desired CVCSs and compare them to the states that would actually be produced by the method, and we show that the latter are in general not known to be useful for measurement-based quantum computation.

The canonical method to generate CVCSs~\cite{Gu2009} is to begin with a collection of highly momentum-squeezed vacuum states (ideally, zero-momentum eigenstates, but these are unphysical) and perform a controlled-Z interaction between neighbouring modes in accord with the desired graph. The family of CVCSs with square-lattice graphs is a universal family of resource states for measurement-based quantum computation using continuous-variable systems~\cite{Menicucci2006}.

A controlled-Z interaction between modes is difficult to implement optically~\cite{Yoshikawa2008}, and as such, more practical methods have been developed to generate CVCSs based on TMS interactions between frequency modes~\cite{Menicucci2007,Zaidi2008,Menicucci2008,Flammia2009,Wang:2013tl} or temporal modes~\cite{Menicucci2011a}. The states generated by these methods are equivalent to CVCSs \emph{only after appropriate phase shifts} have been performed on selected modes~\cite{Menicucci2007,Menicucci2011}.
These multimode-squeezed states (before the phase shifts) are known in the literature as \emph{H-graph states} (for `Hamiltonian graph')~\cite{Menicucci2011} because they are defined in terms of a graph representing a multimode-squeezing Hamiltonian, which, acting on the vacuum, would produce the desired state in the limit of large squeezing~\cite{Menicucci2007,Menicucci2011}. (The state need not actually be made using this H-graph Hamiltonian to be considered an H-graph state.) Such a state has correlations only between like quadratures ($q$-$q$ and $p$-$p$), while CVCSs require $q$-$p$ correlations, which arise after appropriate phase shifts are applied.

Intuitively, one might expect that, since the TMS interactions can be described in graph form (via the H-graph) and since CVCSs have a graph representation as well, there must be a simple relationship between the two graphs. This is not the case in general~\cite{Menicucci2007}. The two graphs are related by the matrix exponential map and additional complicated transformations corresponding to the required phase shifts~\cite{Menicucci2011}, meaning that in general the two graphs bear no resemblance to each other at all~\cite{Menicucci2007}. A simple relationship is only possible under special conditions. One such condition is when the H-graph representing the state is both bipartite and self-inverse (in terms of its adjacency matrix)~\cite{Flammia2009,Menicucci2011}. Very few ordinary graphs (that are sufficiently large) have this property~\cite{Flammia2009}.

Ref.~\cite{Bruschi:2013uv} uses graphs (and graph-like arrow diagrams) to depict the prescribed pattern of TMS interactions but incorrectly claims that they double as the graphs for the resulting CVCSs in general. For example, Fig.~2 of \cite{Bruschi:2013uv} supposedly depicts ``The CVCS generated by the travel scenario described in the text,'' and Fig.~3 of that work has a caption that begins, ``Square cluster state\ldots.'' These are incorrect. They are not (a priori) graphs representing the CVCSs that are generated by the procedures described in the text. Instead, the text indicates that they are simply illustrations of prescribed TMS interactions, acting sequentially and starting with the vacuum, that are implemented by the proposed travel scenarios. Such a sequence of TMS interactions produces an H-graph state, but it still needs to be phase shifted appropriately to transform it into a CVCS of any sort~\cite{Menicucci2011}, and doing so fails, in general, to generate a CVCS with the same structure~\cite{Menicucci2007}. The authors do not take this into account in their proposal and instead simply cite Ref.~\cite{Menicucci2008} while claiming, ``In this work we will exploit particle creation resonances, since the squeezing gates which can be produced by them are useful resource~[sic] for the generation of cluster states.'' No mention is made of the required phase shifts, nor is any attempt made to prove that the resulting CVCS has the same graph as that depicting the TMS interactions.

Furthermore, while the states that result from the TMS interactions are properly called H-graph states (due to same-quadrature correlations), the graphs in Ref.~\cite{Bruschi:2013uv} are not actually the H-graphs for the states because they represent a sequence of noncommuting TMS \emph{unitary interactions} instead of a sum of TMS \emph{Hamiltonians}~\cite{Menicucci2011}. In fact, the graphs used in Ref.~\cite{Bruschi:2013uv} do not correspond to any of the graphs commonly used in the literature (H-graphs or CVCS graphs or Gaussian pure-state graphs~\cite{Menicucci2011}). It is therefore insufficient simply to cite the literature~\cite{Zaidi2008,Menicucci2008,Pysher:2011hn} as justification for the proposal.

\emph{Analysis.}---%
Fortunately, we don't have to rely on the existing literature. We can instead simply follow the procedure as proposed in Ref.~\cite{Bruschi:2013uv} and calculate what state results from it. This is most easily accomplished using the graphical calculus for Gaussian pure states, which provides a visual and mathematically precise description for these states~\cite{Menicucci2011}.

\begin{figure*}[p]
\centering
\begin{tabular}{cc}
(a) & (b) \\
$\vcenter{\hbox{\includegraphics [width=1.2 \columnwidth]{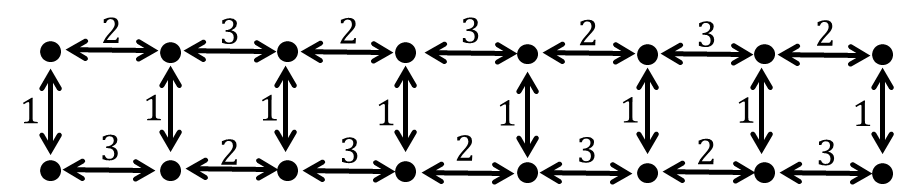}}}$~~~~~~ & ~~~~$\vcenter{\hbox{\includegraphics [width=0.6 \columnwidth]{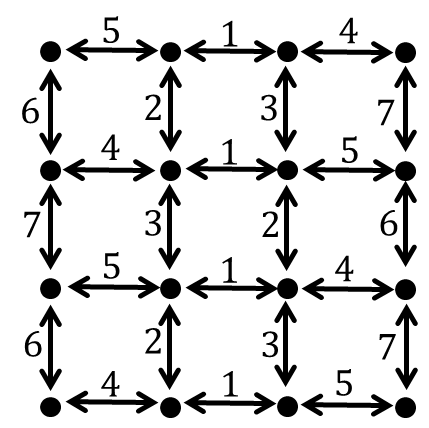}}}$
\end{tabular}
\caption{\label{fig:travelscen}(a)~Identical TMS operations (double-ended arrows) applied to an initial collection of vacuum states (nodes), based on Fig.~2 of Ref.~\cite{Bruschi:2013uv} and the relevant text. The numbers indicate the order in which the interactions are applied (they don't commute in general). (b)~Identical TMS operations applied to initial vacuum states in accord with a square lattice, based on Fig.~3 of Ref.~\cite{Bruschi:2013uv} and the relevant text. Numbers indicate the order.}
\end{figure*}

Ref.~\cite{Bruschi:2013uv} suggests that in order to generate the ``ladder'' state described in Fig.~2 of their paper, a series of TMS interactions are performed on $p-1$ modes by a three-stage travel scenario acting on a collection of vacuum states, where $p$ is prime. We model this procedure by starting with the vacuum, applying three sets of TMS interactions, and calculating the graph corresponding to the resulting Gaussian state. The first interaction is between all modes satisfying the condition $k+k'=p$, the second is between modes satisfying $k+k'=p-2$, and the third is between modes satisfying $k+k'=p+2$. The manner in which these gates are applied is shown in Fig.~\ref{fig:travelscen}(a). The authors claim that this results in a CVCS with the same structure as the ``ladder'' graph used to specify the TMS interactions---Fig.~(2) in Ref.~\cite{Bruschi:2013uv}. %

Choosing $p=17$, we try to generate a 16-mode ladder cluster by applying TMS operations in the manner prescribed. Since the output state is not a CVCS (just an H-graph state, as discussed above), we follow the procedure outlined in Section~IV of Ref.~\cite{Menicucci2011}, which finds the closest CVCS to a given H-graph state (i.e., the CVCS with the least approximation error~\cite{Menicucci2011}) by considering appropriate phase shifts. %
This process produces a Gaussian pure state, and every Gaussian pure state has a corresponding graph~$\mathbf{Z}$, which uniquely identifies it (up to phase-space displacements and overall phase)~\cite{Menicucci2011}. 

An ideal CVCS has a purely real~$\mathbf Z$ corresponding to a useful graph (i.e., one that is connected but still local, e.g., a square lattice), but these states are unphysical~\cite{Menicucci2011}. As such, a good approximation to a useful CVCS is a state whose graph~$\mathbf Z$ satisfies two conditions~\cite{Menicucci2011}: (1)~$\mathbf Z$ has a real part corresponding to a useful graph (e.g., a lattice of some sort), and (2)~$\mathbf Z$ has a very small imaginary part.

Fig.~\ref{fig:ladder}(a) shows the target CVCS: the authors claim that a good approximation to this state will be produced by applying TMS interactions as in Fig.~\ref{fig:travelscen}(a). Figs.~\ref{fig:ladder}(b) and~(c) show the closest CVCS to the actual state produced by this procedure for both low and high squeezing, respectively. Simple visual inspection of these states shows that the target state is not achieved by this method. But let's look more closely at the actual states produced.

Although not really a CVCS at all because it violates condition~(2), we include a low-squeezing example because $\Re(\mathbf Z)$ in~(b) looks approximately like the target state~(a), and we find this interesting. However, the state in this case is completely dominated by $\Im(\mathbf Z)$, which is close to that for a collection of vacuum modes (unsurprising since the squeezing is weak). These states are useless for measurement-based quantum computation because noise will drown out any calculation~\cite{Alexander:MGl69zxv}. %
The high-squeezing limit is, in fact, the only relevant limit if one wants to do measurement-based quantum computation using CVCSs since it is the limit in which H-graph states correspond to CVCSs with high fidelity and since it is the limit in which computations performed using the CVCS may be fault tolerant~\cite{Menicucci:2013vj}. The closest CVCS for high squeezing is shown in~(c) and is approximately a collection of disconnected two-mode pairs (thick lines). Due to its lack of connectivity, this state violates condition~(1) and is therefore not known to be useful for measurement-based quantum computation. (We leave open the possibility that a suboptimal choice of phase shifts may nevertheless produce a useful state with a larger---but still small---approximation error~\cite{Menicucci2011}.) Thus, the proposed construction of ladder-graph CVCSs fails in general.

In Section~III~F of Ref.~\cite{Bruschi:2013uv}, the authors propose extending the ladder graph to a 2D square-lattice structure. This involves repeating the above steps and then implementing further TMS interactions through extra travel scenarios (corresponding to odd numbers $p'$, $p''$, $p'-p+p''-2$ and $3 p -p'-p''+2$). To determine the resulting state, we apply TMS operations to 16 modes (all starting in the vacuum) in accord with the travel scenarios illustrated in Fig.~\ref{fig:travelscen}(b). This results in an H-graph state, which we convert into the nearest CVCS by the procedure described in Ref.~\cite{Menicucci2011}. Fig.~\ref{fig:grid}(a) shows the target square-lattice CVCS. Figs.~\ref{fig:grid}(b) and~(c) show the closest CVCS to the actual state for both low and high squeezing, respectively. The results are similar to those for the ladder: while low squeezing reveals approximately the desired structure, these are useless due to noise (large imaginary part), and the closest CVCS in the high-squeezing limit is just a collection of disconnected two-mode CVCSs, which is not known to be useful due to its disconnected nature. (Once again, we leave open the possibility that suboptimal phase shifts may produce a useful state with suboptimal approximation error~\cite{Menicucci2011}.)

\emph{Conclusion.}---%
we have shown that performing the sequence of TMS operations as described in Ref.~\cite{Bruschi:2013uv} does not in general result in the CVCSs that are claimed. The actual states created are in general not known to be useful for measurement-based quantum computation because the closest CVCS in many cases has a disconnected graph in the high-squeezing limit.

We can think of several possible ways to modify the proposal to produce computationally useful states. First, we note that the authors of Ref.~\cite{Bruschi:2013uv} point out that one can in principle perform arbitrary Bogoliubov transformations using a combination of relativistic motion and natural time evolution of the modes. One possibility would therefore be to more closely mimic existing frequency-mode construction methods~\cite{Menicucci2008,Flammia2009,Wang:2013tl} that use H-graphs of a special form. These provide a simple relationship between the resulting state and closest CVCS.

Second, one might search for ways to use the existing construction with a different prescription for phase shifts, resulting in a CVCS with a larger (but still small) approximation error~\cite{Menicucci2011} but which nevertheless has desirable connectivity in the high-squeezing limit.

Finally, %
one might instead expand the definition of measurement-based quantum computation to allow more general measurements (such as two-mode joint measurements). %
This may be worth exploring if such measurements are more natural in the physical systems considered. In this case, however, the travel scenarios as proposed would almost certainly be unnecessarily complicated since a single travel scenario would, for instance, produce a cleaner copy of a state like Fig.~\ref{fig:ladder}(c) than the three scenarios together as proposed.

\bibliographystyle{bibstyleNCM_papers}
\bibliography{allrefs}

\def\looplessladder{.81}
\def\loopedladder{1.10}

\begin{figure*}[p]
\centering
\begin{tabular}{ccc}
 & $\Re(\mathbf{Z})$ & $i \Im(\mathbf{Z})$ \\[1em]
(a) &
$\vcenter{\hbox{\includegraphics [width=\looplessladder \columnwidth]{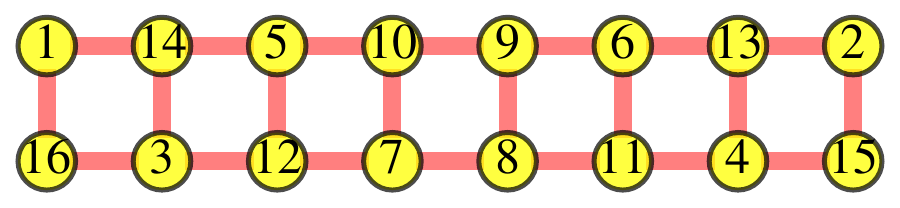}}}$
 &
$\vcenter{\hbox{\includegraphics [width=\looplessladder \columnwidth]{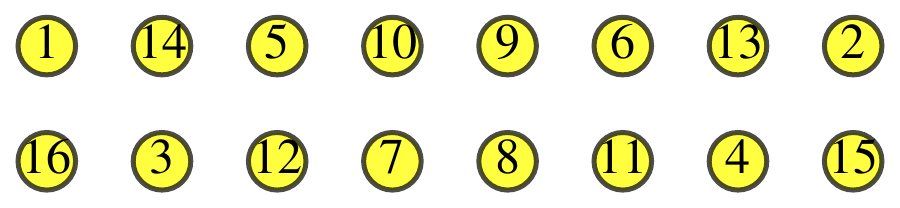}}}$
 \\[3em]
 \hline
(b) &
$\vcenter{\hbox{\includegraphics [width=\looplessladder \columnwidth]{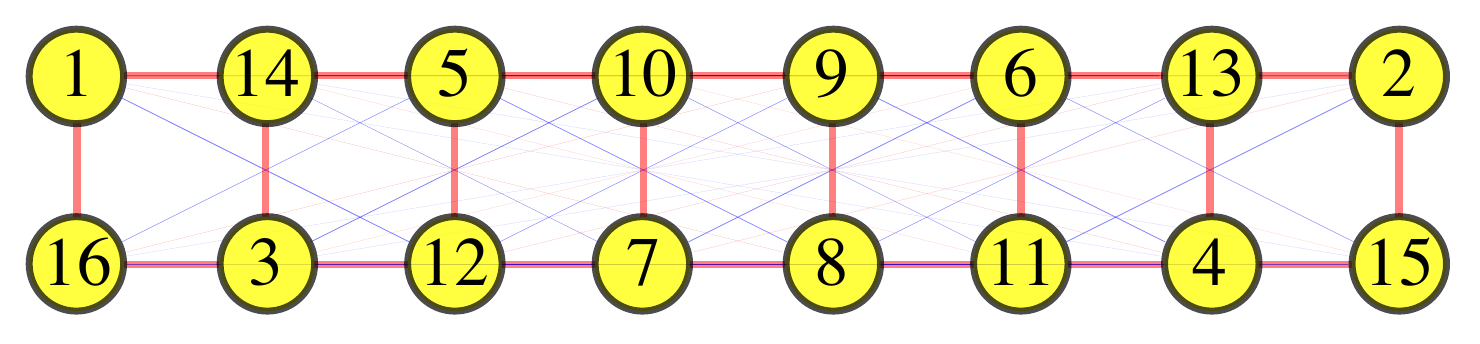}}}$
&
$\vcenter{\hbox{\includegraphics [width=\loopedladder \columnwidth]{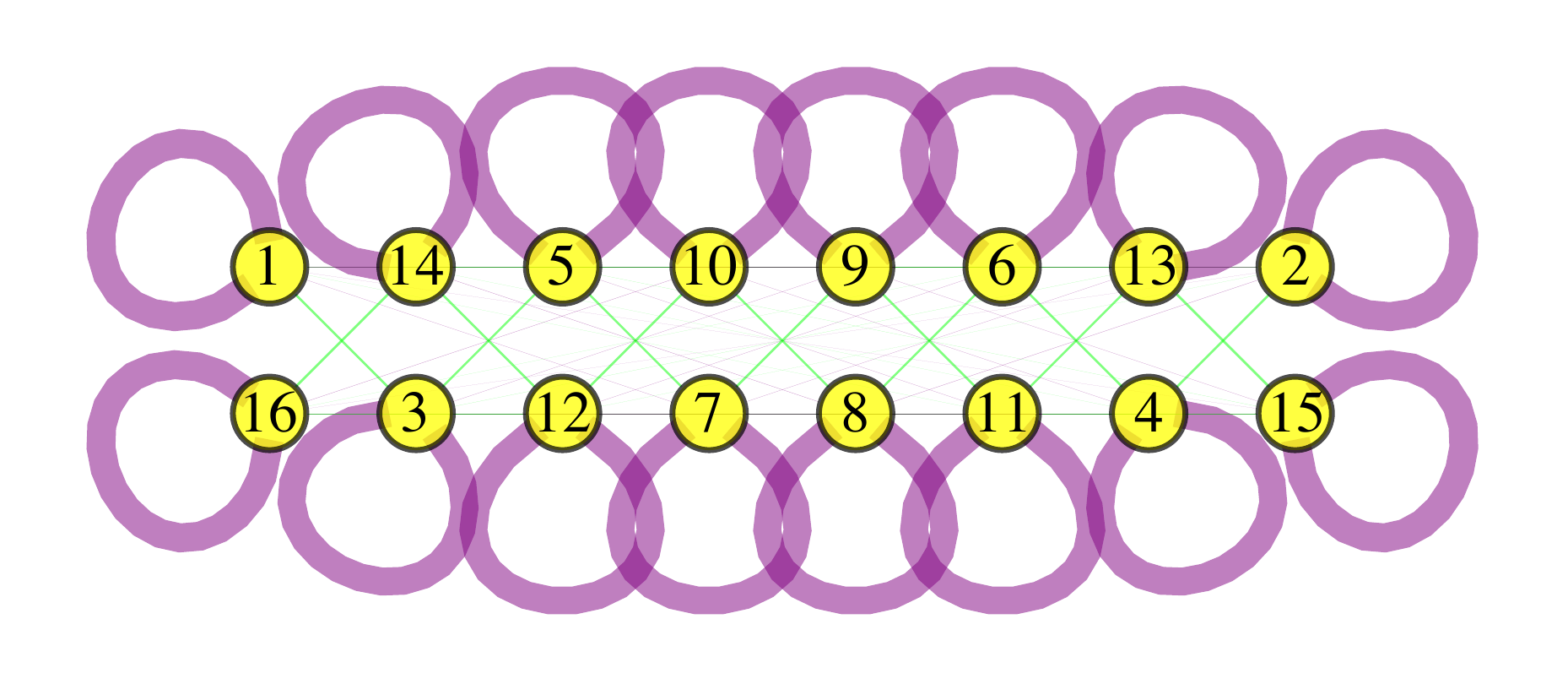}}}$
\\
 \hline(c) &
$\vcenter{\hbox{\includegraphics [width=\looplessladder \columnwidth]{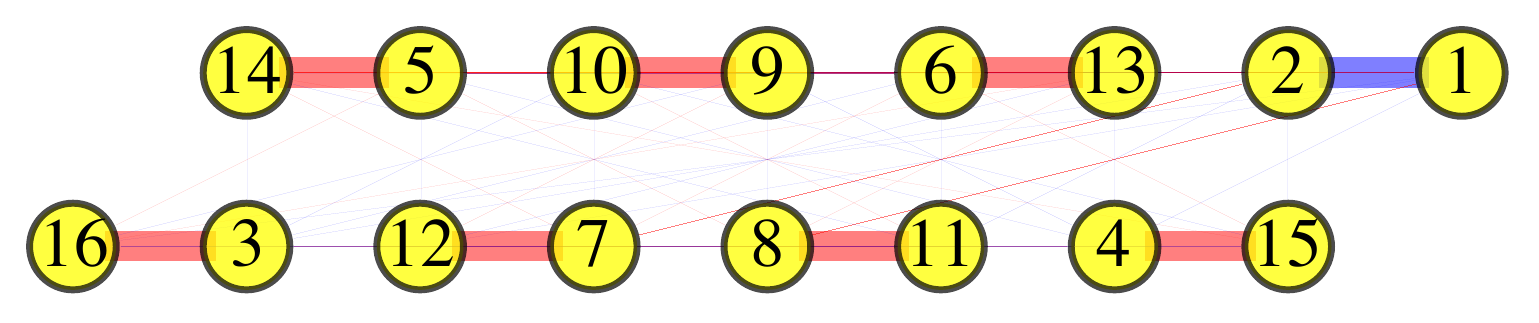}}}$
 &
$\vcenter{\hbox{\includegraphics [width=\loopedladder \columnwidth]{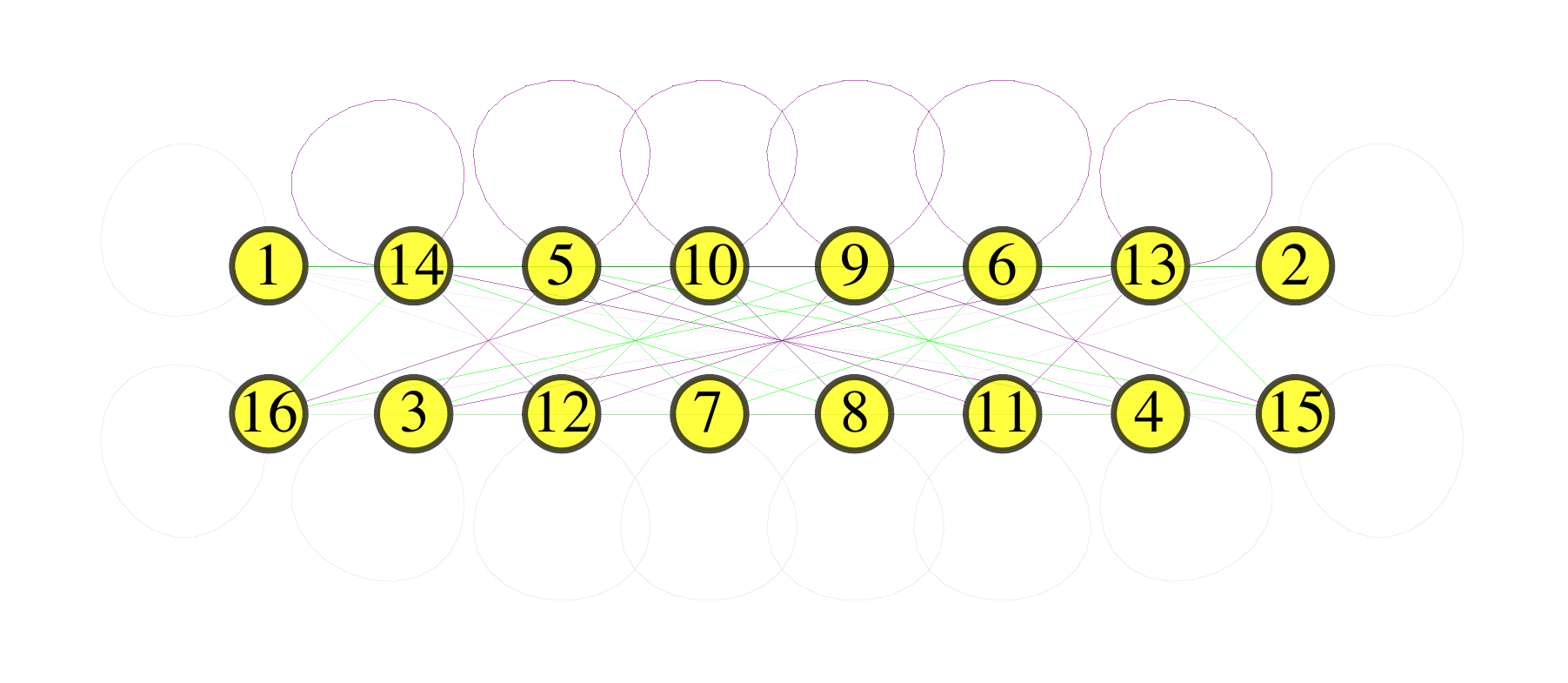}}}$
 \\
\end{tabular}
\caption{\label{fig:ladder}Analysis of the procedure in Fig.~\ref{fig:travelscen}(a), illustrated using the graphical calculus for Gaussian pure states~\cite{Menicucci2011}. The thickness of the graph edges represents the magnitude of the weight of that edge; $(\text{red}, \text{blue}, \text{purple}, \text{green}) \propto (+1, -1, +i, -i)$. (a)~The graph~$\mathbf Z$ for the target continuous-variable cluster state~(CVCS) (with uniform weight~1) claimed in Ref.~\cite{Bruschi:2013uv} to result approximately from this procedure. (b)~The closest CVCS~\cite{Menicucci2011} to the actual state produced for low squeezing---i.e., squeezing parameter~$r = 0.15$ (1.3~dB) for each TMS interaction.  The real part is a ladder with uniform weight~$\approx 0.27$ plus additional edges~$\sim 10^{-2}$. The imaginary part dominates the state, with self loops of weight~$\approx 0.92 i$ and additional edges~$\sim 10^{-1}$. Although $\Re \mathbf Z$ approximates that of~(a) up to an overall constant, since $\Im \mathbf Z \sim \mathbf I$, this is not an approximate CVCS~\cite{Menicucci2011}; high squeezing is needed~\cite{Menicucci2007}. (c)~The closest CVCS for high squeezing, $r = 4.15$ (36~dB) for each TMS interaction. Since the eigenvalues of $\Im \mathbf Z$ are all~$\lesssim 10^{-6}$, the state is an approximate CVCS, but the ideal CVCS it approximates is not the target CVCS in~(a); instead, it approximates 8 disconnected two-mode CVCSs. (The  thick edges have weights~$\approx \pm 1$, with additional edges~$\sim 10^{-2}$.) Its disconnected nature means that this state is not known to be useful for measurement-based quantum computation. The possibility remains that a different choice of phase shifts would result in a useful state but with higher approximation error~\cite{Menicucci2011}.}
\end{figure*}

\def\looplessgrid{.51}
\def\loopedgrid{.80}

\begin{figure*}[p]
\centering
\begin{tabular}{ccc}
 & $\Re(\mathbf{Z})$ & $i \Im(\mathbf{Z})$ \\[1em]
(a) &
$\vcenter{\hbox{\includegraphics [width=\looplessgrid \columnwidth]{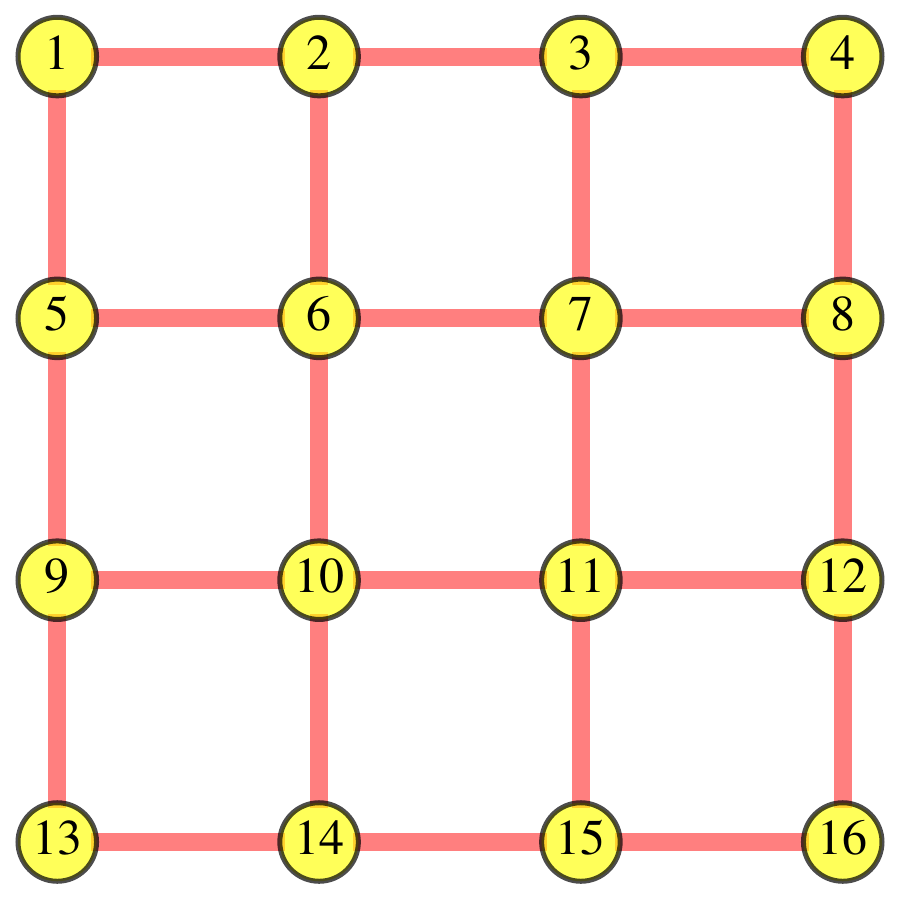}}}$
 &
$\vcenter{\hbox{\includegraphics [width=\looplessgrid \columnwidth]{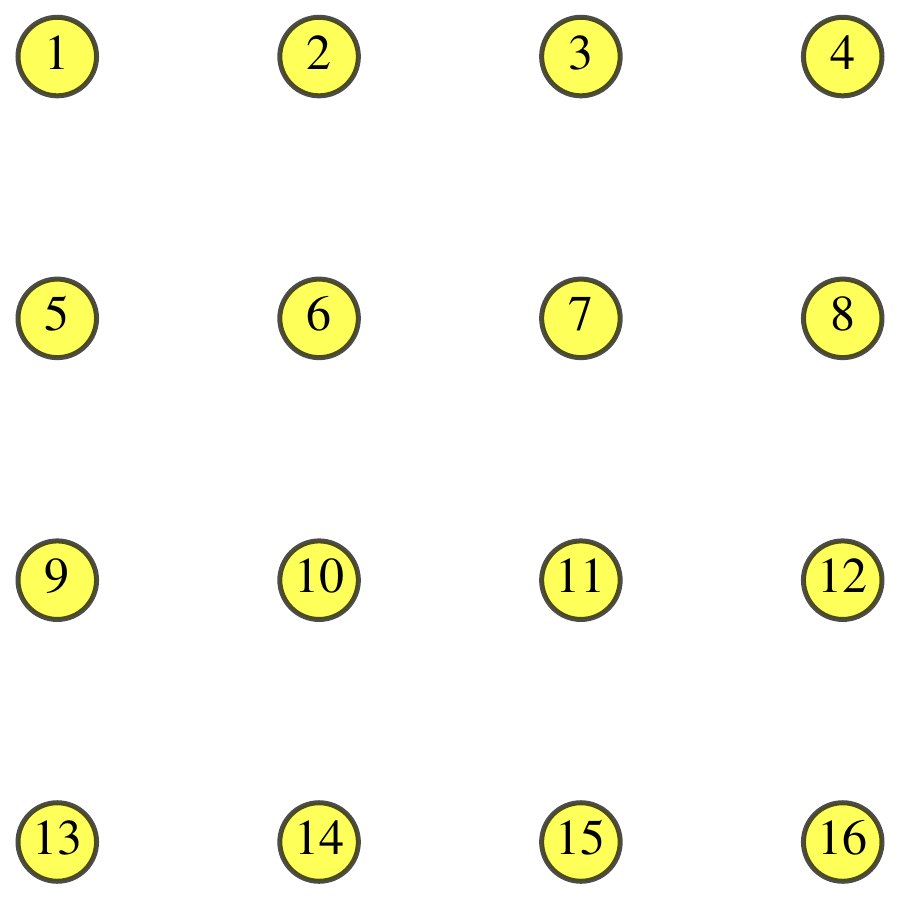}}}$
 \\[7em]
 \hline
(b) &
$\vcenter{\hbox{\includegraphics [width=\looplessgrid \columnwidth]{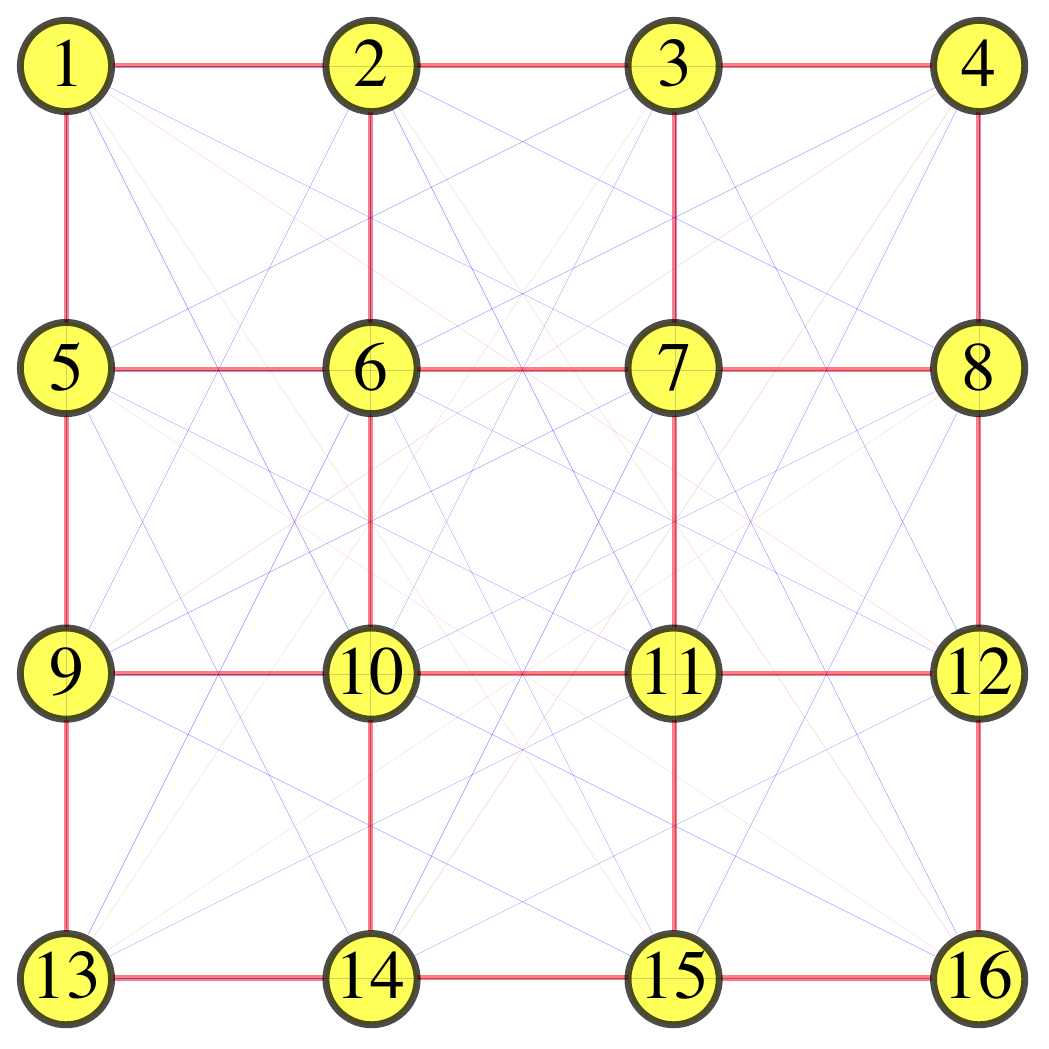}}}$
&
$\vcenter{\hbox{\includegraphics [width=\loopedgrid \columnwidth]{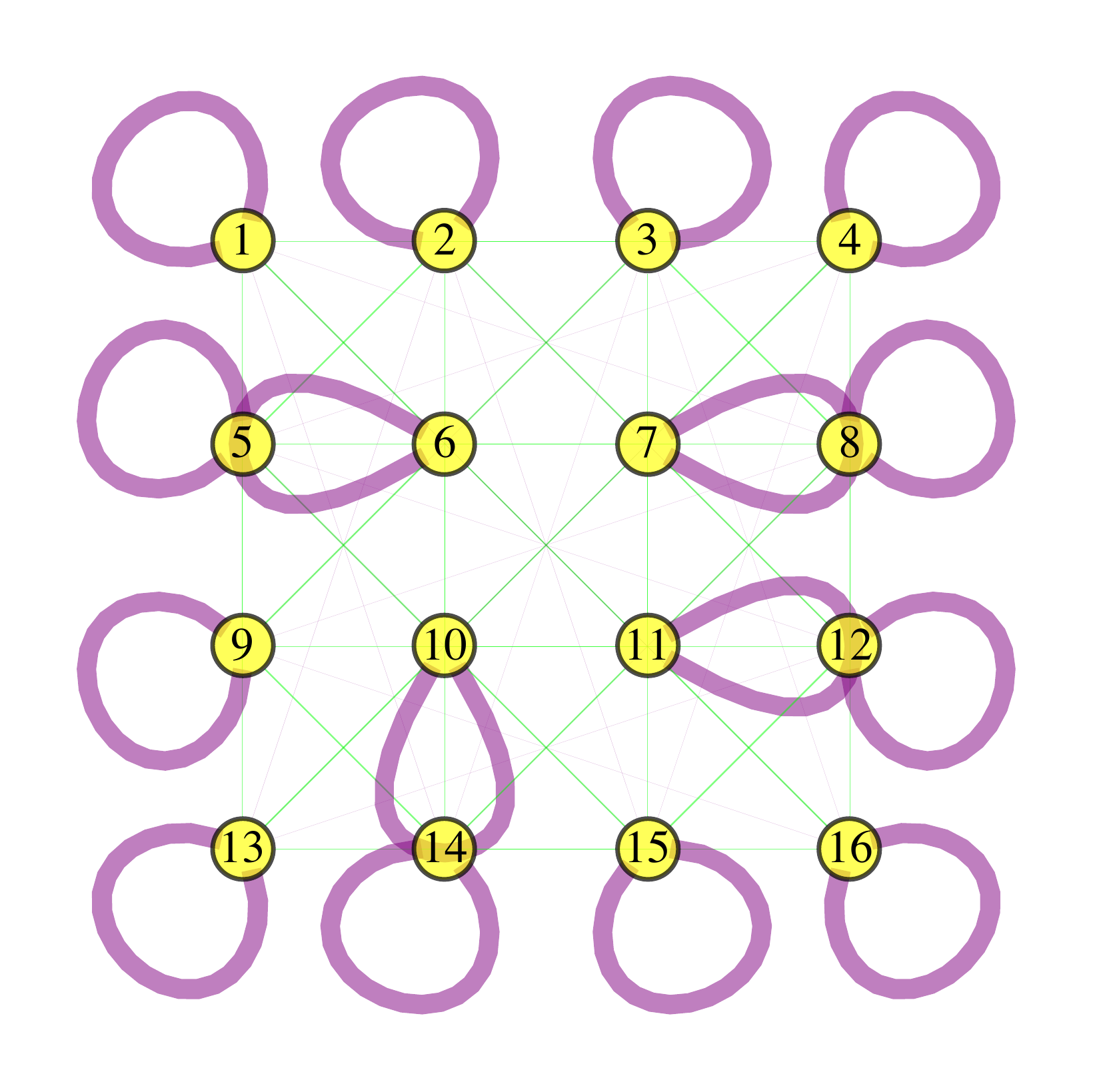}}}$
\\
 \hline
(c) &
$\vcenter{\hbox{\includegraphics [width=\looplessgrid \columnwidth]{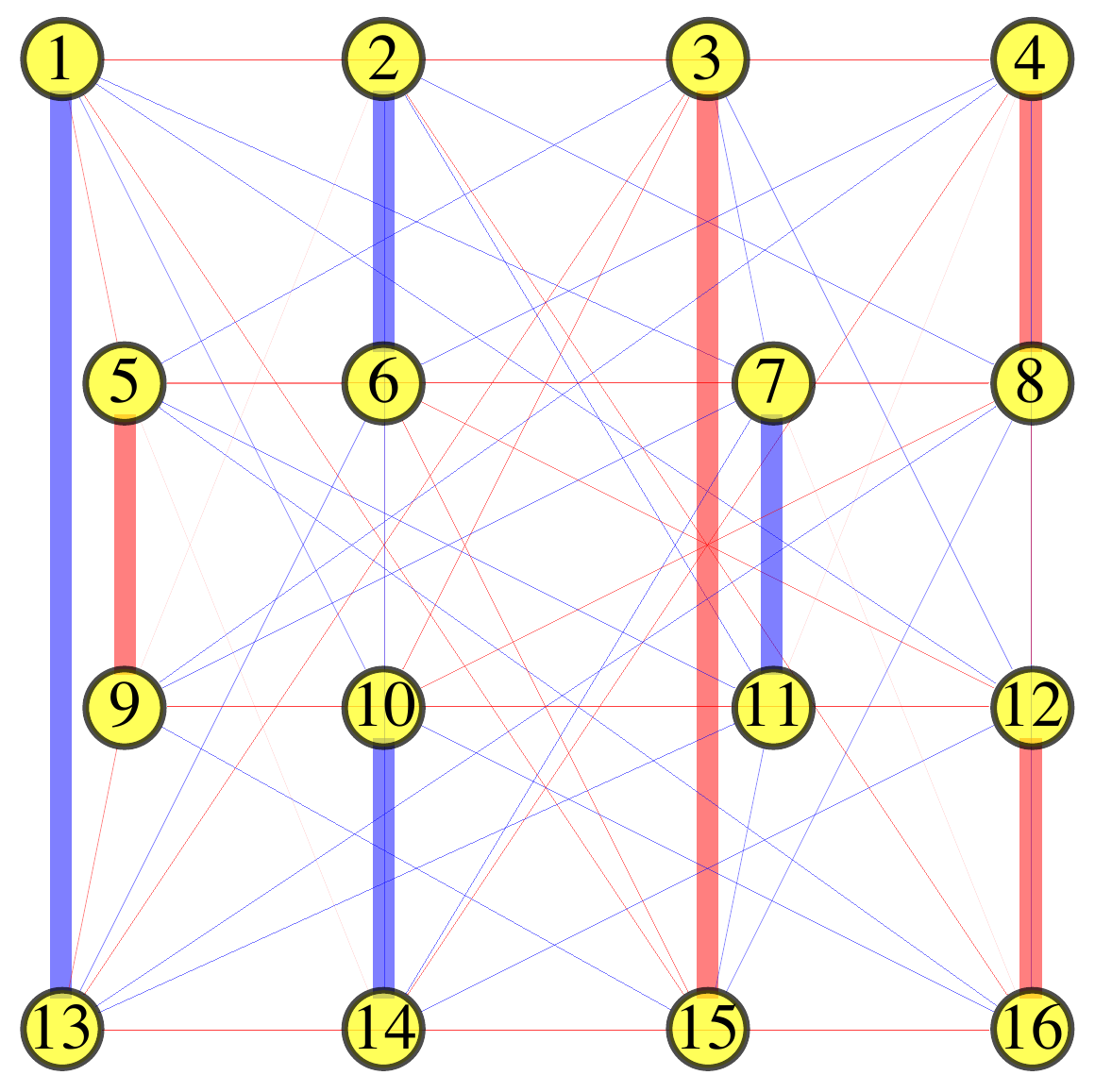}}}$
 &
$\vcenter{\hbox{\includegraphics [width=\loopedgrid \columnwidth]{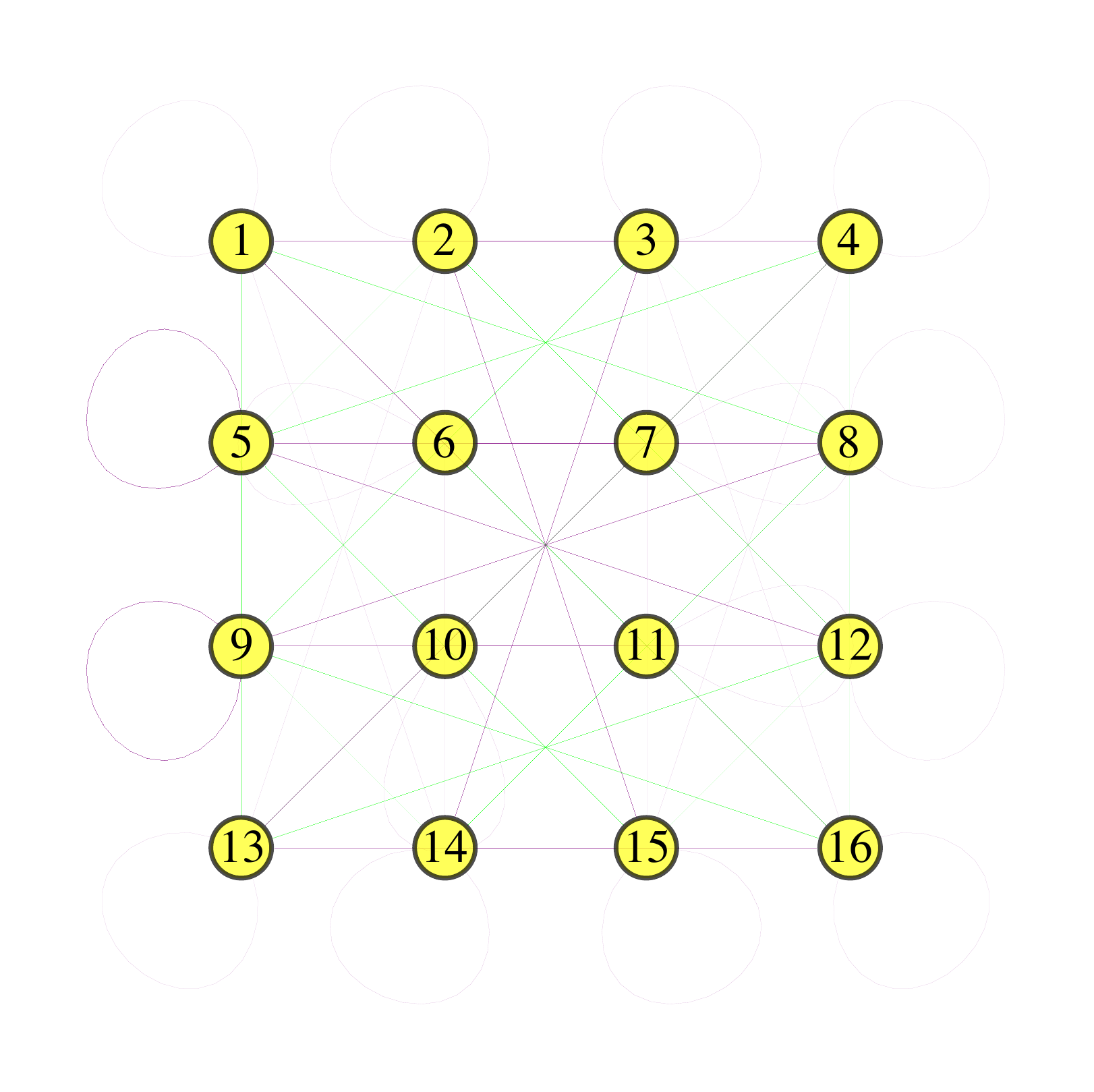}}}$
 \\
\hline
\end{tabular}
\caption{\label{fig:grid}
Analysis of the procedure in Fig.~\ref{fig:travelscen}(b). Same notes on graph representation and squeezing parameters as in Fig.~\ref{fig:ladder}.  %
(a)~The graph~$\mathbf Z$ for the target CVCS (with uniform weight~1) claimed in Ref.~\cite{Bruschi:2013uv} to result approximately from this procedure (arbitrary node numbering). (b)~The closest CVCS~\cite{Menicucci2011} to the actual state produced for low squeezing. The real part is a square lattice with uniform weight~$\approx 0.26$ plus additional edges~$\sim 10^{-2}$. The imaginary part dominates, with self loops of weight~$\approx 0.92 i$ and additional edges~$\sim 10^{-1}$. Although $\Re \mathbf Z$ approximates that of~(a) up to an overall constant, since $\Im \mathbf Z \sim \mathbf I$, this is not an approximate CVCS~\cite{Menicucci2011}; high squeezing is needed~\cite{Menicucci2007}. (c)~The closest CVCS for high squeezing. Since the eigenvalues of $\Im \mathbf Z$ are all~$\lesssim 10^{-3}$, the state is an approximate CVCS, but the ideal CVCS it approximates is not the target CVCS in~(a); instead, it approximates 8 disconnected two-mode CVCSs. (The  thick edges have weights~$\approx \pm 1$, with additional edges~$\sim 10^{-2}$.) Its disconnected nature means that this state is not known to be useful for measurement-based quantum computation. The possibility remains that a different choice of phase shifts would result in a useful state but with higher approximation error~\cite{Menicucci2011}.
}
\end{figure*}

\end{document}

%% file: macros.tex
\usepackage{color}
\usepackage{amsmath,amsfonts,amssymb,amsthm}
\usepackage{graphicx}
\PassOptionsToPackage{caption=false}{subfig}

\usepackage{tikz}
\usetikzlibrary{arrows,decorations.pathmorphing,backgrounds,positioning,fit}
\usepackage{epstopdf} %
\usepackage{bm}  %

\usepackage[pdfpagelabels,pdftex,bookmarks,breaklinks]{hyperref}
\definecolor{darkblue}{RGB}{0,0,127} %
\definecolor{darkgreen}{RGB}{0,150,0}
\hypersetup{colorlinks, linkcolor=darkblue, citecolor=darkgreen, filecolor=red, urlcolor=blue}

\usepackage{layout}

\pgfrealjobname{RelativisticCVCSs}

\makeatletter
\@ifundefined{textcolor}{}
{%
 \definecolor{BLACK}{gray}{0}
 \definecolor{WHITE}{gray}{1}
 \definecolor{RED}{rgb}{1,0,0}
 \definecolor{GREEN}{rgb}{0,.4,0}
 \definecolor{BLUE}{rgb}{0,0,1}
 \definecolor{CYAN}{cmyk}{1,0,0,0}
 \definecolor{MAGENTA}{cmyk}{0,1,0,0}
 \definecolor{YELLOW}{cmyk}{.2,.4,1,0}
 }
\makeatother

\renewcommand{\Re}{\operatorname{Re}}
\renewcommand{\Im}{\operatorname{Im}}

%% file: RelativisticCVCSs.bbl
\begin{thebibliography}{10}

\bibitem{Bruschi:2013uv}
D.~E. Bruschi, C. Sab{\'\i}n, P. Kok, G. Johansson, P. Delsing, and I. Fuentes,
  ``{Towards universal quantum computation through relativistic motion},''
  arxiv:1311.5619v1 [quant-ph] (2013).

\bibitem{Gu2009}
M. Gu, C. Weedbrook, N.~C. Menicucci, T.~C. Ralph, and P. van Loock, ``{Quantum
  Computing with Continuous-Variable Clusters},'' Phys. Rev. A {\bf 79}, 062318
  (2009).

\bibitem{Menicucci2006}
N.~C. Menicucci, P. van Loock, M. Gu, C. Weedbrook, T.~C. Ralph, and M.~A.
  Nielsen, ``{Universal Quantum Computation with Continuous-Variable Cluster
  States},'' Phys. Rev. Lett. {\bf 97}, 110501 (2006).

\bibitem{Yoshikawa2008}
J.-i. Yoshikawa, Y. Miwa, A. Huck, U.~L. Andersen, P. van Loock, and A.
  Furusawa, ``{Demonstration of a quantum nondemolition sum gate},'' Phys. Rev.
  Lett. {\bf 101}, 250501 (2008).

\bibitem{Menicucci2007}
N.~C. Menicucci, S.~T. Flammia, H. Zaidi, and O. Pfister, ``{Ultracompact
  generation of continuous-variable cluster states},'' Phys. Rev. A {\bf 76},
  010302(R) (2007).

\bibitem{Zaidi2008}
H. Zaidi, N.~C. Menicucci, S.~T. Flammia, R. Bloomer, M. Pysher, and O.
  Pfister, ``{Entangling the optical frequency comb: simultaneous generation of
  multiple 2x2 and 2x3 continuous-variable cluster states in a single optical
  parametric oscillator},'' Laser Phys. {\bf 18}, 659 (2008).

\bibitem{Menicucci2008}
N.~C. Menicucci, S.~T. Flammia, and O. Pfister, ``{One-Way Quantum Computing in
  the Optical Frequency Comb},'' Phys. Rev. Lett. {\bf 101}, 130501 (2008).

\bibitem{Flammia2009}
S.~T. Flammia, N.~C. Menicucci, and O. Pfister, ``{The Optical Frequency Comb
  as a One-Way Quantum Computer},'' J. Phys. B {\bf 42}, 114009 (2009).

\bibitem{Wang:2013tl}
P. Wang, M. Chen, N.~C. Menicucci, and O. Pfister, ``{Weaving quantum optical
  frequency combs into hypercubic cluster states},'' arxiv:1309.4105v1
  [quant-ph] (2013).

\bibitem{Menicucci2011a}
N.~C. Menicucci, ``{Temporal-mode continuous-variable cluster states using
  linear optics},'' Phys. Rev. A {\bf 83}, 062314 (2011).

\bibitem{Menicucci2011}
N.~C. Menicucci, S.~T. Flammia, and P. van Loock, ``{Graphical calculus for
  Gaussian pure states},'' Phys. Rev. A {\bf 83}, 042335 (2011).

\bibitem{Pysher:2011hn}
M. Pysher, Y. Miwa, R. Shahrokhshahi, R. Bloomer, and O. Pfister, ``{Parallel
  Generation of Quadripartite Cluster Entanglement in the Optical Frequency
  Comb},'' Phys. Rev. Lett. {\bf 107}, 030505 (2011).

\bibitem{Alexander:MGl69zxv}
R.~N. Alexander, S.~C. Armstrong, R. Ukai, and N.~C. Menicucci, ``{Noise
  analysis of single-qumode Gaussian operations using continuous-variable
  cluster states},'' arxiv:1311.3538v1 [quant-ph] (2013).

\bibitem{Menicucci:2013vj}
N.~C. Menicucci, ``{Fault-tolerant measurement-based quantum computing with
  continuous-variable cluster states},'' arxiv:1310.7596v2 [quant-ph] (2013).

\end{thebibliography}
